\documentclass[12pt,a4paper]{article}
\usepackage{amsmath} 
\usepackage{amssymb}
\usepackage{graphicx,epsfig}
\usepackage[mathscr]{eucal}
\usepackage{color} 
\usepackage[numbers,sort&compress]{natbib}
\usepackage{setspace}
\usepackage{xspace}

\usepackage{multirow}

\newcommand\as{\alpha_{\mathrm{S}}}

\def\to{\rightarrow} 
\def\nn{\nonumber}

\def\ms{\ensuremath{{\overline {\rm MS}}}\xspace} 
\newcommand\OS{{OS}\xspace}

\def\nnlosv{\ensuremath{{\rm NNLO}_{\rm SV}}\xspace} 
\def\nlosv{\ensuremath{{\rm NLO}_{\rm SV}}\xspace}

\def\Mt{\ensuremath{{M_{t}}}\xspace}

\def\mmunew{m_t(\mu_m)}
\def\lmnew{{L_{\mmunew} } }
\def\muR{\ensuremath{\mu^{}_R}\xspace}
\def\muF{\ensuremath{\mu^{}_F}\xspace}
\def\mum{\ensuremath{\mu^{}_m}\xspace}
\def\muRsq{\ensuremath{\mu^{2}_R}\xspace}

\def\mumsq{\ensuremath{\mu^{2}_m}\xspace}
\def\z#1{\zeta_{#1}}

\usepackage{array}
\newcolumntype{L}[1]{>{\raggedright\let\newline\\\arraybackslash\hspace{0pt}}m{#1}}
\newcolumntype{C}[1]{>{\centering\let\newline\\\arraybackslash\hspace{0pt}}m{#1}}
\newcolumntype{R}[1]{>{\raggedleft\let\newline\\\arraybackslash\hspace{0pt}}m{#1}}

\usepackage[hidelinks]{hyperref}
\usepackage{footnotebackref}

\begin{document} 
\begin{titlepage}
\begin{flushright}
MPP-2022-63
\end{flushright}

\renewcommand{\thefootnote}{\fnsymbol{footnote}}
\vspace*{0.2cm}

\begin{center}
  {\Large \bf
  NNLO study of top-quark mass renormalization   \\[0.5cm]
   scheme uncertainties in Higgs boson production}
\end{center}

\par \vspace{2mm}
\begin{center}
  {\bf Javier Mazzitelli}

\vspace{5mm}

Max-Planck Institut f\"ur Physik, F\"ohringer Ring 6, 80805 M\"unchen, Germany

\vspace{5mm}

\end{center}

\par \vspace{1mm}
\begin{center} {\large \bf Abstract} 

\end{center}
\begin{quote}
\pretolerance 10000

The ambiguity in the choice of a renormalization scheme and scale for the top-quark mass leads to an additional source of theoretical uncertainty in the calculation of the Higgs boson production cross section via gluon fusion. These uncertainties were found to be dominant in the case of off-shell Higgs production at next-to-leading order in QCD for large values of the Higgs virtuality $m_H^*$.
In this work, we study the uncertainties related to the top-quark mass definition up to next-to-next-to-leading order (NNLO) in QCD.
We include the full top-quark mass dependence up to three loops in the virtual corrections, and evaluate the real contributions in the soft limit, therefore obtaining the so-called soft-virtual (SV) approximation.
We construct \nnlosv predictions for off-shell Higgs boson production renormalizing the top-quark mass within both the on-shell (OS) and the \ms schemes, and study in detail the differences between them.
While the differences between the two schemes are sizeable, we find that the predictions are always compatible within scale uncertainties. We also observe that the difference between renormalization schemes is largely reduced when increasing the order of the perturbative expansion.
We analyze the quality of the convergence of the perturbative series in both schemes, and find that at large invariant masses the \ms results present much larger corrections than their \OS counterparts.
We also comment on the more complicated case of Higgs boson pair production.

\end{quote}

\vspace*{\fill}
\begin{flushleft}
June 2022
\end{flushleft}
\end{titlepage}

\section{Introduction}
\label{sec:intro}

A decade after the discovery of the Higgs boson~\cite{Englert:1964et,Higgs:1964ia,Higgs:1964pj,Guralnik:1964eu} at the Large Hadron Collider (LHC)~\cite{ATLAS:2012yve,CMS:2012qbp}, the study of its properties is still one of the main quests of the particle physics community.
The ATLAS and CMS collaborations have a rich program for present and future runs devoted to the exploration of the Higgs sector, and these searches are not limited to the production and decay of an on-shell Higgs boson.
In fact, the production of an off-shell Higgs boson is crucial, for instance, for imposing constraints on its decay width~\cite{Caola:2013yja,ATLAS:2018jym,CMS:2019ekd}. Furthermore, the decay of an off-shell Higgs boson into pairs provides a handle on the Higgs self coupling, which is explored through the di-Higgs production process (see ref.~\cite{DiMicco:2019ngk} for a review). In both cases, the main production mode at the LHC is gluon fusion, mediated by a heavy-quark loop.

The experimental searches mentioned above rely on accurate theoretical predictions, and to this end leading-order (LO) calculations in the QCD perturbative expansion are completely inadequate, and the effect from higher-order corrections needs to be included. The next-to-LO (NLO) corrections for Higgs boson production with full top-quark mass dependence are known since a long time~\cite{Graudenz:1992pv,Spira:1995rr}, however the complete next-to-NLO (NNLO) predictions have only been computed recently~\cite{Czakon:2021yub}, for the production of an on-shell Higgs boson with mass $m_H=125$~GeV.
The QCD corrections have also been evaluated in the heavy top-quark mass limit, with the current state-of-the-art being the next-to-NNLO (N$^3$LO)~\cite{Anastasiou:2015vya,Mistlberger:2018etf}, together with partial higher-order results~\cite{Bonvini:2014joa,Schmidt:2015cea,Das:2020adl,Ajjath:2021bbm}.
The residual theoretical uncertainties due to the truncation of the QCD perturbative expansion are estimated via scale variations, and found to be about $3\%$ at N$^3$LO.

In the calculation of the Higgs boson production cross section via gluon fusion, the renormalization of the top-quark mass is a scheme-dependent procedure (see Ref.~\cite{Saibel:2021krs} for a NLO study on $t\bar{t}H$ production).
The most common choices are the on-shell (\OS) and the \ms schemes. In the former, the pole of the top-quark propagator is fixed to the same value at any order in the perturbative expansion; said value is the so-called pole mass $M_t$. In the latter, only the singular pieces (working in dimensional regularization) of the ultraviolet divergencies are removed, and the pole of the quark propagator receives corrections at any order of the perturbative expansion. The \ms mass $m_t(\mu_m)$ therefore differs from $M_t$, and in addition depends on the renormalization scale $\mu_m$.

The ambiguity in the choice of a scheme and a scale for the renormalization of the top-quark mass leads to ensuing uncertainties. These are found to be sub-leading for the production of an on-shell Higgs boson, with the differences between the cross sections in the \OS and \ms schemes being at the per-mille level at NLO.\footnote{Note that here and throughout this work we are only considering the top-quark contribution to the Higgs production cross section, and the uncertainties related to its renormalization scheme and scale. Contributions from lighter quarks and their scheme uncertainties are not addressed in this work.}
However, when considering the production of an off-shell Higgs boson (similar considerations apply to the production of Higgs boson pairs), the larger scale $m_H^*>m_H$ leads to a stronger parametric dependence on the top-quark mass, and consequently to a larger scheme and scale dependence.
This dependence was studied at NLO in refs.~\cite{Baglio:2018lrj,Baglio:2020ini,Amoroso:2020lgh} and found to be very large, typically exceeding the size of other theoretical uncertainties for large values of $m_H^*$, and being dominant as well in the case of Higgs boson pair production, both in the total cross section and in the invariant mass ($m_{HH}$) distribution.
In order to control this source of uncertainties, higher-order studies of the top-quark mass scheme and scale dependence are needed.

In this work, we present an NNLO study of the scheme and scale uncertainties for off-shell Higgs boson production via gluon fusion. Based on the three-loop corrections with full top-quark mass dependence presented in ref.~\cite{Czakon:2020vql}, we compute the NNLO cross section for Higgs boson production in the soft-virtual approximation. We then consider both the \OS and the \ms schemes, obtaining numerical predictions for the LHC up to NNLO. Our results represent the first NNLO-accurate study of this source of uncertainty.

This paper is organized as follows: in section~\ref{sec:calculation} we provide the technical details of our calculation. In section~\ref{sec:results} we present the phenomenological results, and compare the results obtained using different schemes and scales for the renormalization of the top-quark mass. We also comment on the case of Higgs boson pair production. Finally, in section~\ref{sec:conclusions} we present our conclusions.

\section{Technical aspects of the calculation}
\label{sec:calculation}

We will consider the total cross section for the production of an off-shell Higgs boson in gluon fusion, mediated via a top-quark loop.
Since we aim to study the uncertainties arising from the definition of the top-quark mass, the full dependence on this parameter needs to be included, and we cannot rely on predictions computed in the heavy-top limit.
Up to NLO, the results needed can be obtained via the public code \verb|iHixs|~\cite{Dulat:2018rbf} (more specifically, \verb|iHixs| can compute the cross section corresponding to a heavy Higgs boson, which is related to the off-shell SM cross section by the corresponding Higgs boson propagator). The code also allows to compute predictions both in the \OS as in the \ms scheme, including the full top-quark mass dependence up to two loops. Similar results can also be obtained with the public code \verb|HIGLU|~\cite{Spira:1996if}.

Predictions at NNLO with full top-quark mass dependence have been obtained in ref.~\cite{Czakon:2021yub}. Those results, however, are limited to the production of an on-shell Higgs boson with mass $m_H=125$~GeV.
In order to obtain NNLO predictions with full top-quark mass dependence and for arbitrary values of $m_H^*$, we rely on the three-loop virtual corrections obtained in ref.~\cite{Czakon:2020vql}. Based on them and on the results of ref.~\cite{deFlorian:2012za}, we construct the soft-virtual approximation at NNLO (\nnlosv).

Denoting by $s$ the hadronic centre-of-mass energy, and by $x_{1,2}$ the momentum fractions of the colliding partons, the partonic centre-of-mass energy is given by $\hat{s} = x_1 x_2 s$.
Given that the parton densities (and especially the gluon densities) grow fast for small momentum fractions, the amount of energy available in the production of a heavy final state will typically be close to the minimum needed to produce such state, predominantly only allowing for additional soft radiation.
In other words, if we consider the variable $z=(m_H^*)^2/\hat{s}$, the dominant contributions to the total cross section arise from the $z\sim 1$ region, that is, the soft region.
More specifically, in the soft-virtual approximation the partonic cross section only contains terms proportional to $\delta(1-z)$ and to the customary plus distributions ${\cal D}_i(z)$, while terms that are regular in the $z \to 1$ limit are neglected.
Note that the virtual corrections, which do not contain any additional real radiation, only contribute to the $\delta(1-z)$ term. Soft emissions, on the other hand, contribute to both $\delta(1-z)$ and ${\cal D}_i(z)$ terms.
We also point out that only the flavour-diagonal channel (in the case of Higgs production, only the gluon-initiated channel) contributes to the cross section in the soft-virtual limit.

Technically speaking, we define the soft-virtual approximation by considering the Mellin transform of the partonic cross section. In that case, the $z\to 1$ limit corresponds to the large-$N$ limit, where $N$ is the conjugate variable of $z$. The soft-virtual cross section is then defined by dropping all the terms that vanish in the limit $N\to\infty$. More details about the exact definition can be found in ref.~\cite{deFlorian:2012za}.

From the computational point of view, the soft-virtual approximation can be obtained by replacing the full real radiation matrix elements by their corresponding soft limit. In this limit, the phase-space integrals become much simpler, and closed expressions can be obtained. In fact, a fully general result for the production of a colour singlet in hadron collisions at \nnlosv has been obtained in ref.~\cite{deFlorian:2012za} (and extended to the next perturbative order in ref.~\cite{Catani:2014uta}). The only process-dependent piece of the final result is given by the finite remainders of the NLO and NNLO virtual corrections.
In other words, with the results of ref.~\cite{deFlorian:2012za} as a starting point, the virtual corrections in ref.~\cite{Czakon:2020vql} are the only ingredient needed to obtain \nnlosv predictions for Higgs boson production.
We stress that when computing predictions at \nnlosv, we always keep the full NLO corrections, and the soft-virtual approximation is only performed in the ${\cal O}(\as^4)$ term.
We also note that \nnlosv results for Higgs boson production were first obtained in the heavy-top limit in ref.~\cite{Catani:2001ic}.

With the implementation described above, we are able to obtain \nnlosv predictions for arbitrary values of $m_H^*$ and $M_t$ in the \OS scheme. For our purposes, we still need to perform the conversion to the \ms scheme, which is briefly discussed below.

The top-quark mass in the \OS scheme, $M_t$, and the corresponding \ms definition, $m(\mu_m)$, are connected by the following relation:
\begin{equation}
\label{eq:polemsbar}
\Mt = m_t(\mu_m) \,d(m_t(\mu_m), \mu_m) =
m_t(\mu_m) \left(
1+\sum_{k=1}^\infty  \left(\frac{\as(\mu_m)}{\pi}\right)^k\, d^{(k)}(\mu_m)
\right).
\end{equation}
The first two perturbative coefficients $d^{(1)}$ and $d^{(2)}$ in eq.~(\ref{eq:polemsbar}) have the values~\cite{Gray:1990yh,Fleischer:1998dw}
\begin{align}
  d^{(1)}(\mu_m) &= \frac{4}{3} + \lmnew\nonumber \,,
  \\[1ex]
  d^{(2)}(\mu_m) &= \frac{307}{32} + 2 \* \z2 + \frac{2}{3} \* \z2 \* \ln 2 - \frac{1}{6}\*\z3 
  + \frac{509}{72}\*\lmnew
  + \frac{47}{24}\*\lmnew^2 
  \nonumber\\
  &- \left( \frac{71}{144} + \frac{1}{3}\*\z2 + \frac{13}{36}\*\lmnew + \frac{1}{12}\*\lmnew^2 \right)\*n_f 
  \, ,
\label{eq:dcoef}
\end{align}
where $n_f=5$ is the number of light quark flavours and
\begin{equation}
  \lmnew = 2 \ln(\mu_m/\mmunew) \,.
\end{equation}
The cross section in the \ms scheme, $\bar\sigma$, can be obtained from the prediction in the on-shell scheme, $\sigma$, through the following formal replacement:
\begin{equation}
\label{eq:all}
\bar\sigma(m_t(\mu_m);\mu_m,\muR,\muF)
= \sigma(\Mt=m_t(\mu_m)\,d(m_t(\mu_m),\mu_m);\muR,\muF) \;.
\end{equation}
In eq.~(\ref{eq:all}), the pole mass $M_t$ in the \OS prediction (right hand side) is expressed in terms of the \ms mass using eq.~(\ref{eq:polemsbar}).
Note that, following eq.~(\ref{eq:all}), $\bar{\sigma}$ and $\sigma$ are identical if considered at all-orders. However, their perturbative expansion will differ order-by-order.

By performing the perturbative expansion of the right hand side of eq.~(\ref{eq:all}), we arrive at the following result for the NNLO prediction in the \ms scheme:
\begin{equation}
  \label{eq:xs_MSbar}
  \bar\sigma_\text{NNLO}(m_t(\mum);\mu_m,\muR,\muF) = 
    \sum_{i=0}^2 \! \left(\frac{\as(\muR)}{\pi}\right)^{i+2} \bar\sigma^{(i)}(m_t(\mum);\mu_m,\muR,\muF) 
  \, ,
\end{equation}
where the purely NNLO contribution $\bar\sigma^{(2)}$ is given by~\cite{Catani:2020tko}
\begin{eqnarray}
 \label{eq:barnnlo}
\bar\sigma^{(2)}(m_t(\mum);\mu_m,\muR,\muF) &=&
   \Bigg[ \sigma^{(2)}(m;\muR,\muF)  \\
 && \hspace*{-3cm}
 + \; m \Bigg(
   d^{(1)}(\mum) \,\partial_{m} \sigma^{(1)}(m;\muR,\muF) 
   + \frac{1}{2} \left(d^{(1)}(\mum)\right)^2 \,m\,
       \partial^2_{m} \sigma^{(0)}(m;\muF) \nn\\ 
 && \hspace*{-3cm}
       + \; d^{(2)}(\mum) \,\partial_{m} \sigma^{(0)}(m;\muF)
       + \beta_0 \, d^{(1)}(\mum) \ln\left(\frac{\muRsq}{\mumsq} \right)     
     \partial_{m} \sigma^{(0)}(m;\muF)
   \Bigg)
   \Bigg]_{m=m_t(\mum)} \;,\nn
\end{eqnarray}
with $\beta_0=\frac{11 C_A}{12} - \frac{n_f}{6}$.
Therefore, the NNLO cross section in the \ms scheme can be obtained from the corresponding \OS prediction evaluated at $M_t=m_t(\mum)$, plus additional terms proportional to the derivatives of the lower-order \OS cross section w.r.t. the top-quark mass. Following the approach of ref.~\cite{Catani:2020tko}, those derivatives are computed numerically by repeating the calculation for several mass values around $M_t=m_t(\mum)$. We have checked that our approach to compute the \ms cross section perfectly agrees with the \ms predictions that can be obtained from \verb|iHixs| at NLO.

\section{Phenomenological results}
\label{sec:results}

In the following, we present our phenomenological results for hadronic collisions at 13~TeV. We consider the range of invariant masses $m_H^* \in (200\text{ GeV},\,1200\text{ GeV})$. Regarding the top-quark mass, we use the value $M_t=172.5$~GeV for the \OS-scheme predictions, while the value $m_t(m_t)=162.9$~GeV is used in the \ms scheme. These two values are related by the conversion relation in eq.~(\ref{eq:polemsbar}) at three-loop order. We use the \verb|PDF4LHC15_nnlo_mc|~\cite{Butterworth:2015oua} parton distribution functions in all our predictions, with the corresponding strong coupling constant.
In the figures where uncertainty bands are shown, we choose a central scale $\mu_0=m_H^*/2$, and we consider independent variations of the scales $\mu_{m,R,F}=\xi\, \mu_0$ with $\xi=\{1/2,1,2\}$, in addition constraining the ratio between any two scales to be not larger than 2. This procedure corresponds to the customary 7-point variation in the \OS scheme, while in the case of the \ms predictions, where one additional scale is present, it leads to a 15-point variation, as introduced in ref.~\cite{Catani:2020tko} in the context of top-quark pair production.

\begin{figure}
\begin{center}
\includegraphics[width=0.49\textwidth]{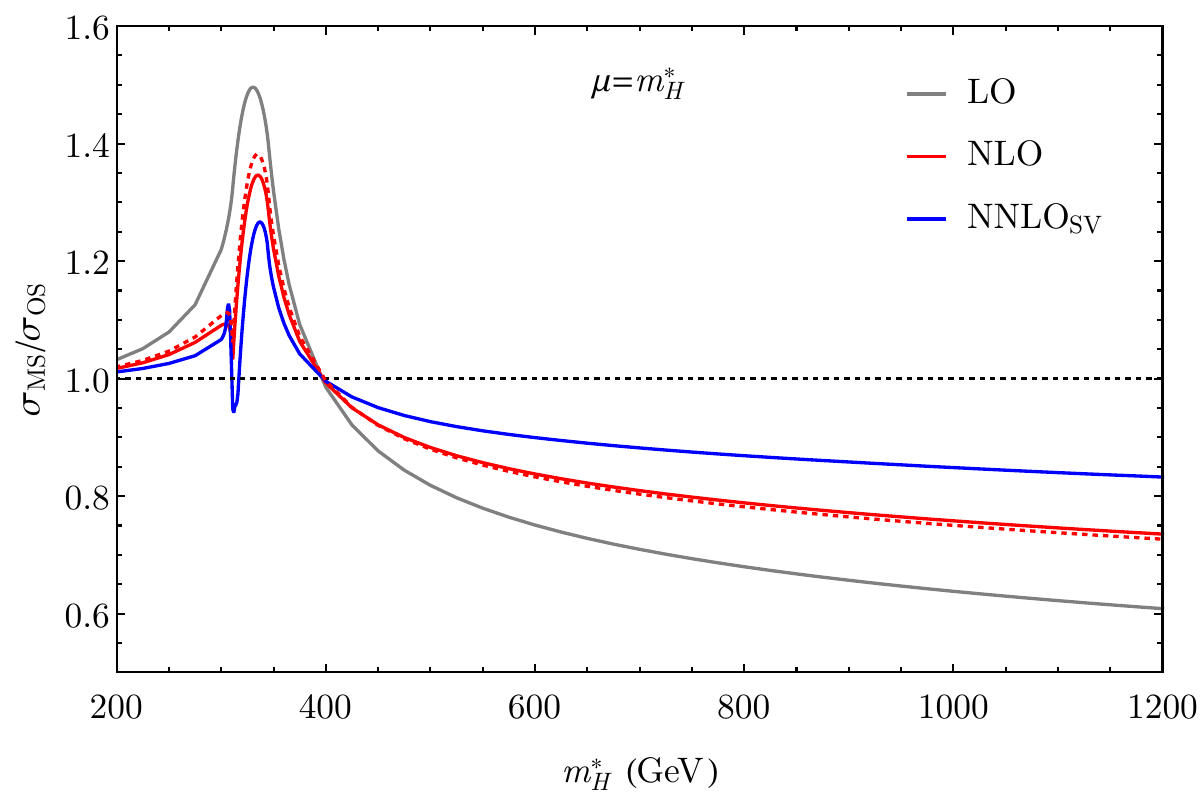}
\includegraphics[width=0.49\textwidth]{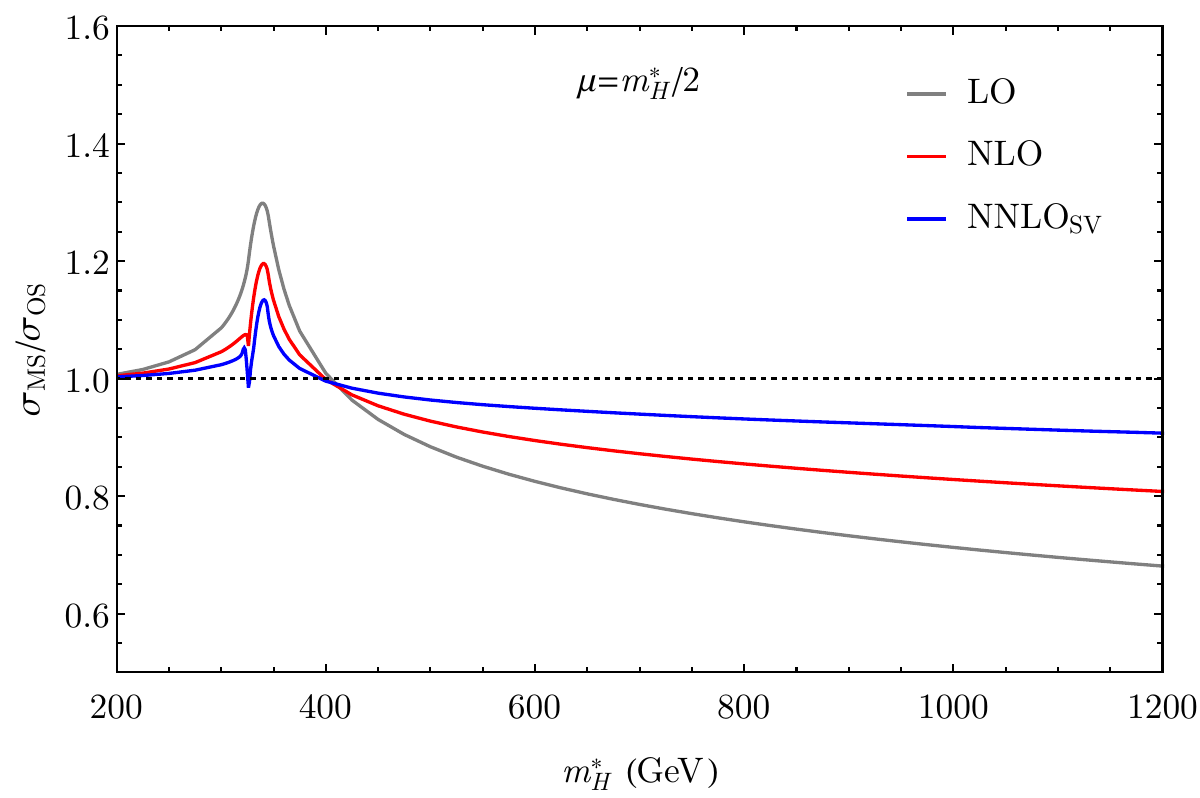}
\end{center}
\vspace{-4ex}
\caption{\label{fig:MS_over_OS}\small
Ratio of the off-shell Higgs production cross section in the \ms scheme to the corresponding \OS prediction, as a function of $m_H^*$. The ratio is computed at LO (gray), NLO (red) and \nnlosv (blue), for a scale $\mu=m_H^*$ (left) and $\mu=m_H^*/2$ (right). In the left panel the \nlosv prediction is also shown (red-dotted line).
}
\end{figure}

We are interested in studying the differences between the \OS and \ms predictions. Therefore, in figure~\ref{fig:MS_over_OS} we present the ratio between the \ms and \OS results, $\bar{\sigma}_{\rm{N}^i\rm{LO}}/\sigma_{\rm{N}^i\rm{LO}}$, as a function of the invariant mass $m_H^*$.
For the sake of clarity, we do not include in this figure uncertainty bands, but rather present results for fixed values of the renormalization and factorization scales.
The results in the left (right) panel correspond to $\mum=\muR=\muF=\mu$ with $\mu=m_H^*$ ($\mu=m_H^*/2$). Note that the same perturbative order and scale choice is used in the numerator and denominator for the construction of each curve.

We first discuss about the validity of the soft-virtual approximation. To that end, we present as well \nlosv predictions (red-dotted line in figure~\ref{fig:MS_over_OS}, left panel) which can be compared to the results obtained with the full NLO corrections. As can be seen from figure~\ref{fig:MS_over_OS}, the \nlosv results reproduce the full NLO curve with a very high accuracy, in particular much higher than the size of the effects we are aiming to assess (that is, the difference between \ms and \OS results). The difference between the NLO and \nlosv curves in figure~\ref{fig:MS_over_OS} is below 1\% for most of the distribution, with a larger deviation found only around the peak that is however always below 3\%.
We therefore consider the \nnlosv results to be an excellent proxy for the full NNLO prediction in the context of this study.

We now focus on the comparison between the two schemes at the different perturbative orders. In the first place, we can observe that in the whole invariant mass region and for both scale choices the differences between the two schemes are reduced from LO to NLO, and from NLO to \nnlosv as well. The gap between the \OS and \ms predictions is larger for $\mu=m_H^*$, which is expected since a higher scale generates a smaller value of $m_t(\mu)$, and therefore a larger difference w.r.t. $M_t$.
In the region $m_H^*<400$~GeV, the \ms scheme predicts a larger cross section. The maximum deviation is about 50\% at LO for $\mu=m_H^*$, going down to 35\% and 27\% at NLO and \nnlosv, respectively. In the case of $\mu=m_H^*/2$, the peak difference between \ms and \OS predictions is 30\%, 20\% and 13\% at LO, NLO and \nnlosv, respectively, in the same invariant mass region.
We observe that the \ms predictions at NLO and \nnlosv present large variations close to the top-quark mass threshold; this can be traced back to the large values of the derivatives w.r.t. the mass that enter their calculation. 

At large invariant mass values, $m_H^*>400$~GeV, the \ms prediction instead provides a lower cross section than its \OS counterpart. The difference between the two schemes steadily increases with $m_H^*$, reaching up to $-39\%$ at LO for $m_H^*=1200$~GeV and $\mu=m_H^*$, with this difference decreasing to $-26\%$ at NLO, and further down to $-17\%$ at \nnlosv. Once again, the predictions for $\mu=m_H^*/2$ present smaller differences, with LO, NLO and \nnlosv respectively reaching a difference of $-32\%$, $-19\%$ and $-9\%$ w.r.t. the \OS prediction.

\begin{figure}
\begin{center}
\includegraphics[width=0.31\textwidth]{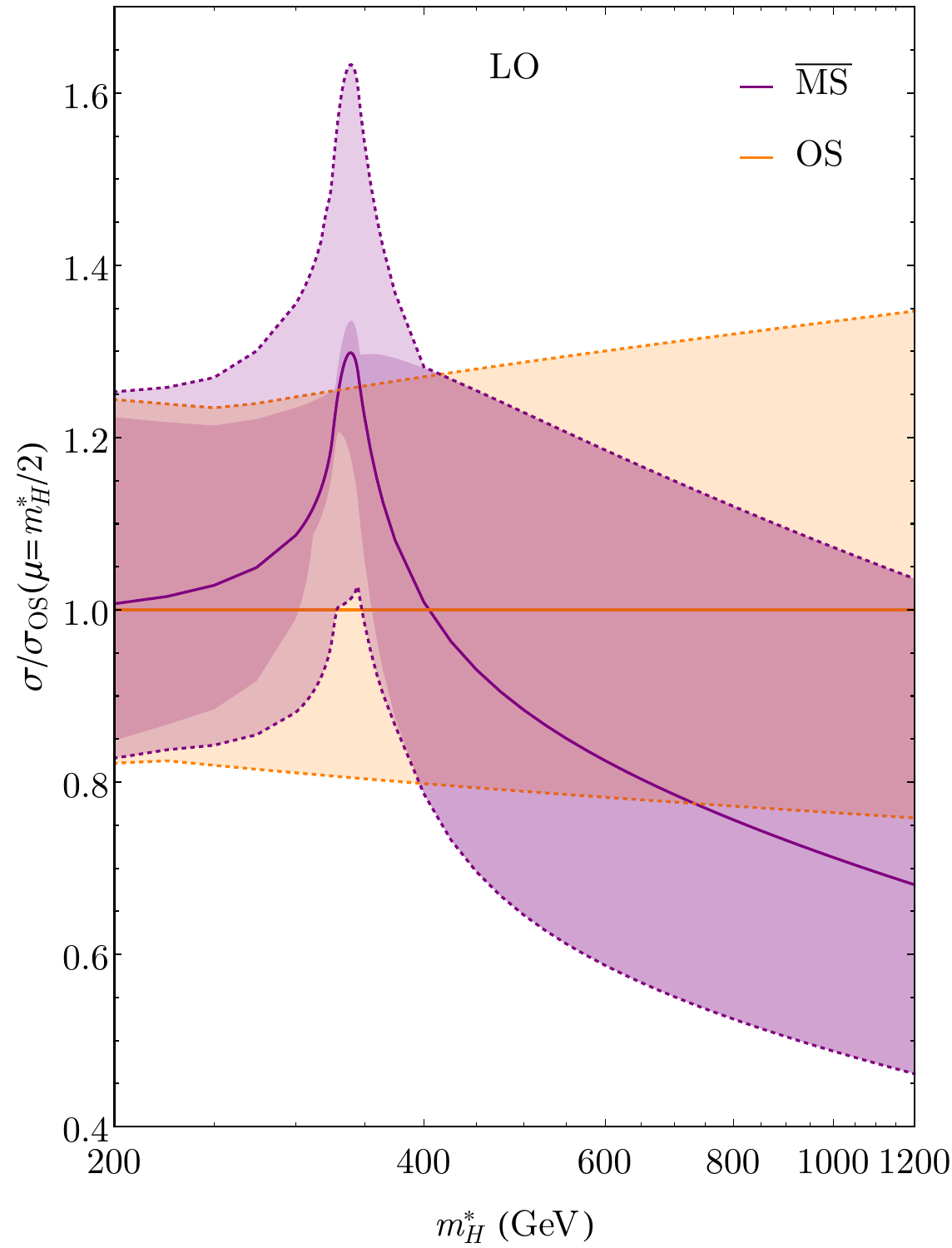}
\includegraphics[width=0.31\textwidth]{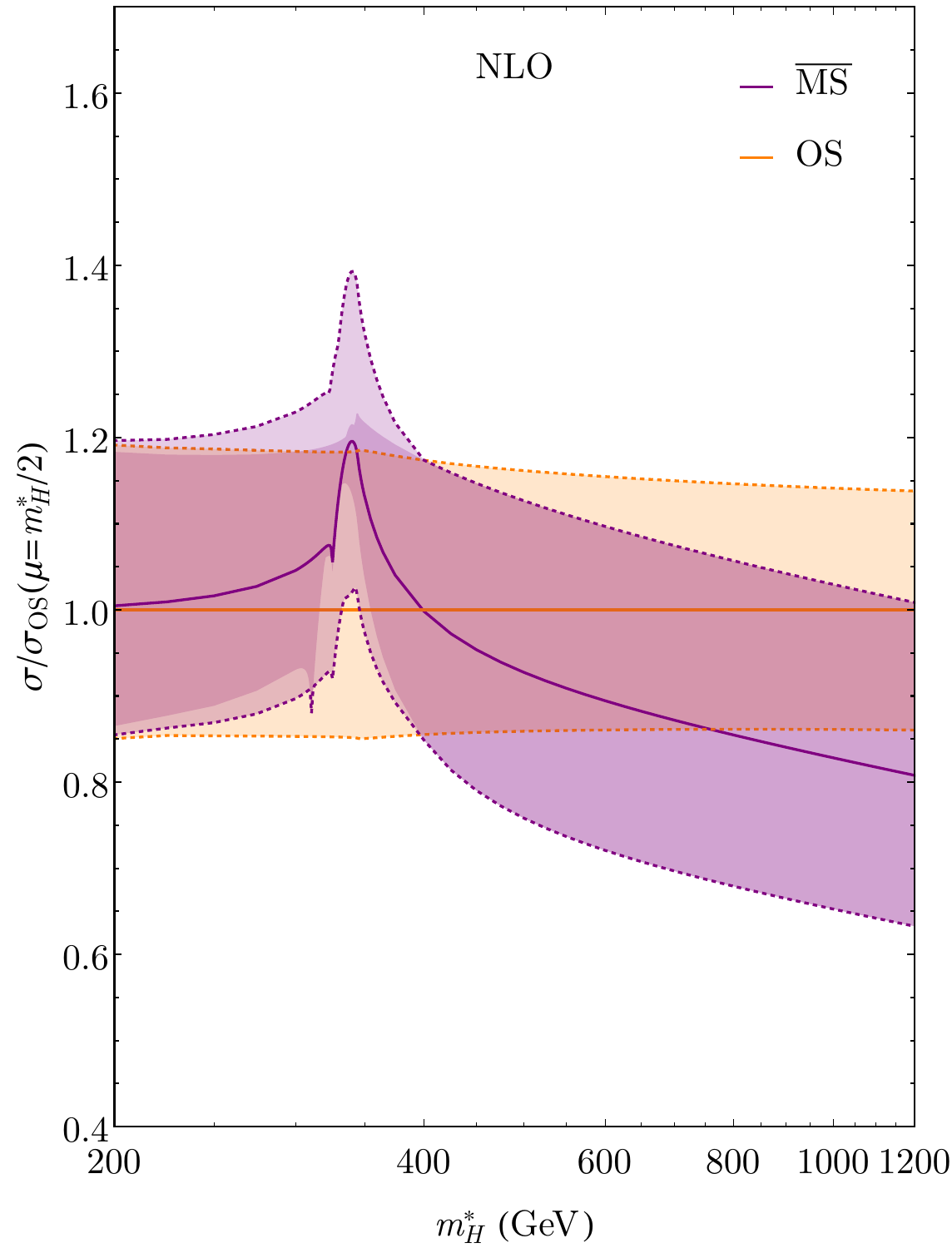}
\includegraphics[width=0.31\textwidth]{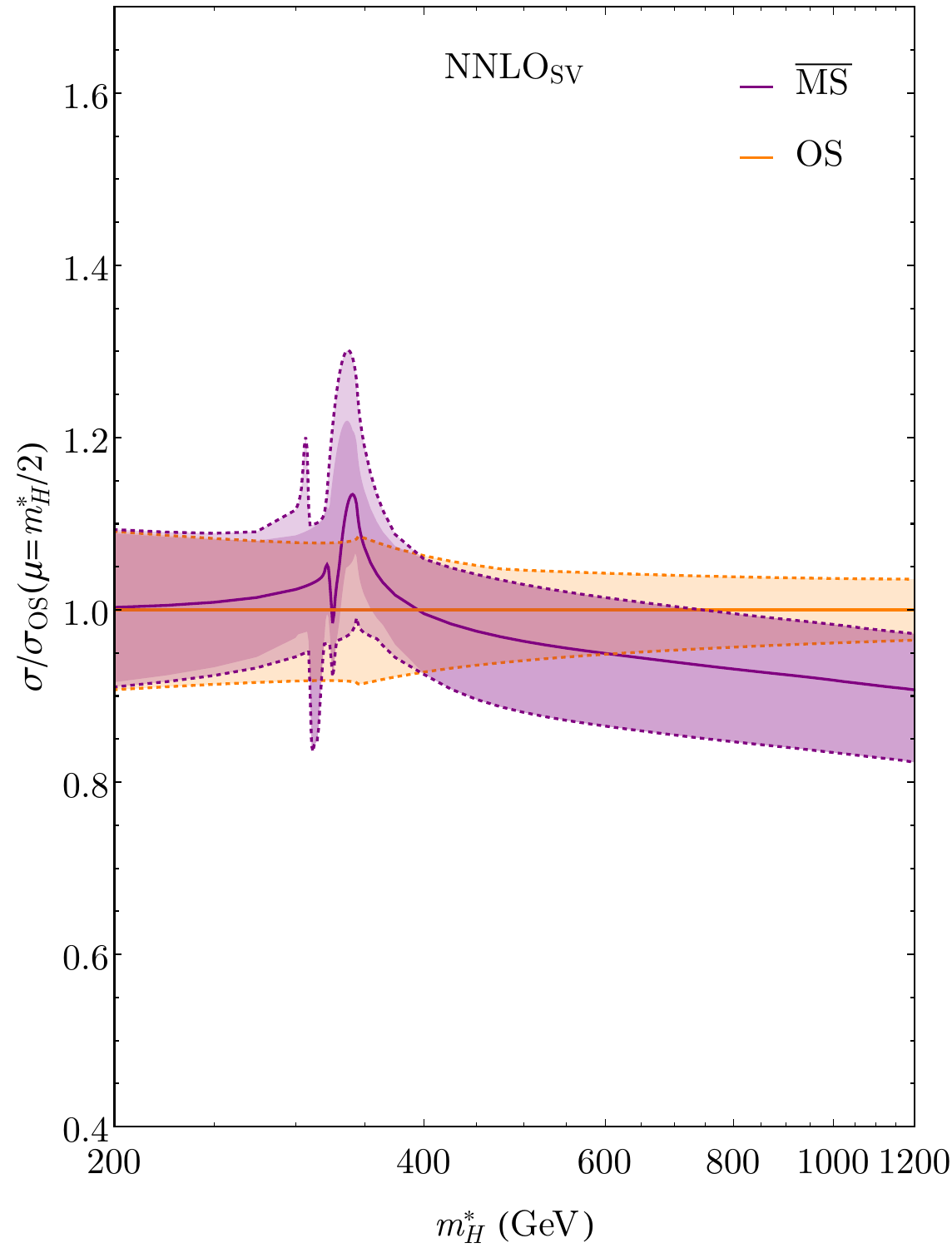}
\end{center}
\vspace{-4ex}
\caption{\label{fig:MS_over_OS_bands}\small
Ratio of the \ms (purple) and \OS (orange) cross sections at LO (left), NLO (center) and \nnlosv (right). The bands denote the scale uncertainties, computed using 7-point and 15-point variations in the \OS and \ms schemes, respectively. The \ms uncertainties corresponding to a restricted 7-point variation with the contraint $\mum=\muR$ are indicated with the darker purple bands.
}
\end{figure}

We study next the same ratio, this time including the scale variation as defined before, and considering $\mu_0=m_H^*/2$ as the central scale. This is shown in figure~\ref{fig:MS_over_OS_bands}, where the different panels (left, centre, right) correspond to the different perturbative orders (LO, NLO, \nnlosv). In this case, the denominator is always computed using $\mu_{m,R,F}=\mu_0$.

In the first place, we can observe that the scale uncertainties in both schemes are significantly reduced when increasing the order of the perturbative expansion. There is always a sizeable overlap between the results of both schemes at each order, indicating that, at least up to NNLO, scale variations are a reliable tool to estimate the theoretical uncertainties arising from the ambiguity in the top-quark mass definition. We note, however, that at \nnlosv and large values of $m_H^*$ the overlap between the \OS and \ms predictions is marginal.

When looking at the general behaviour of the curves presented in figure~\ref{fig:MS_over_OS_bands}, we can see three different regions with qualitatively different features. For low values of $m_H^*$, both schemes yield similar results, as expected due to the reduced parametric dependence on $M_t$ in said region. This changes dramatically as we cross the $t\bar{t}$ threshold, and a further change in the behaviour is observed for large values of the Higgs invariant mass.
We address these last two regions in more detail in the following paragraphs.

As mentioned before, the region close to the $t\bar{t}$ threshold presents a large dependence on the renormalization-scheme choice, due to the strong dependence of the cross section on $M_t$. The production rate is increased when crossing this threshold, and the large variations observed originate from the shift in the exact position of this enhancement when changing the value of the top-quark mass.

In the case of on-shell $t\bar{t}$ production, it has been shown that the use of the \ms scheme leads to large theoretical uncertainties close to threshold~\cite{Dowling:2013baa,Catani:2020tko}. For a proper estimation of these uncertainties, it was crucial to consider independent variations of the top-quark renormalization scale $\mum$~\cite{Catani:2020tko}.
We can observe that the latter also occurs in the present case of off-shell Higgs boson production. To illustrate this, in figure~\ref{fig:MS_over_OS_bands} we show the \ms scale uncertainties within the full 15-point variation, and also the restricted 7-point variation corresponding to the constraint $\mum=\muR$ (darker band).

As can be observed from the plot, the restricted 7-point variation would lead to a large underestimation of the scale uncertainties. This situation is specific of this invariant mass region, where the size of the variation induced by \mum and \muR in the cross section is of similar magnitude, but opposite sign. In contrast, for values of $m_H^*$ much lower than $2M_t$ the dependence on \mum is much smaller, and therefore the scale variation is driven by \muR, while for invariant masses above the $t\bar{t}$ threshold the \mum and \muR variations have the same sign (though, again, they are mostly driven by \muR in this region).

\begin{figure}
\begin{center}
\includegraphics[width=0.49\textwidth]{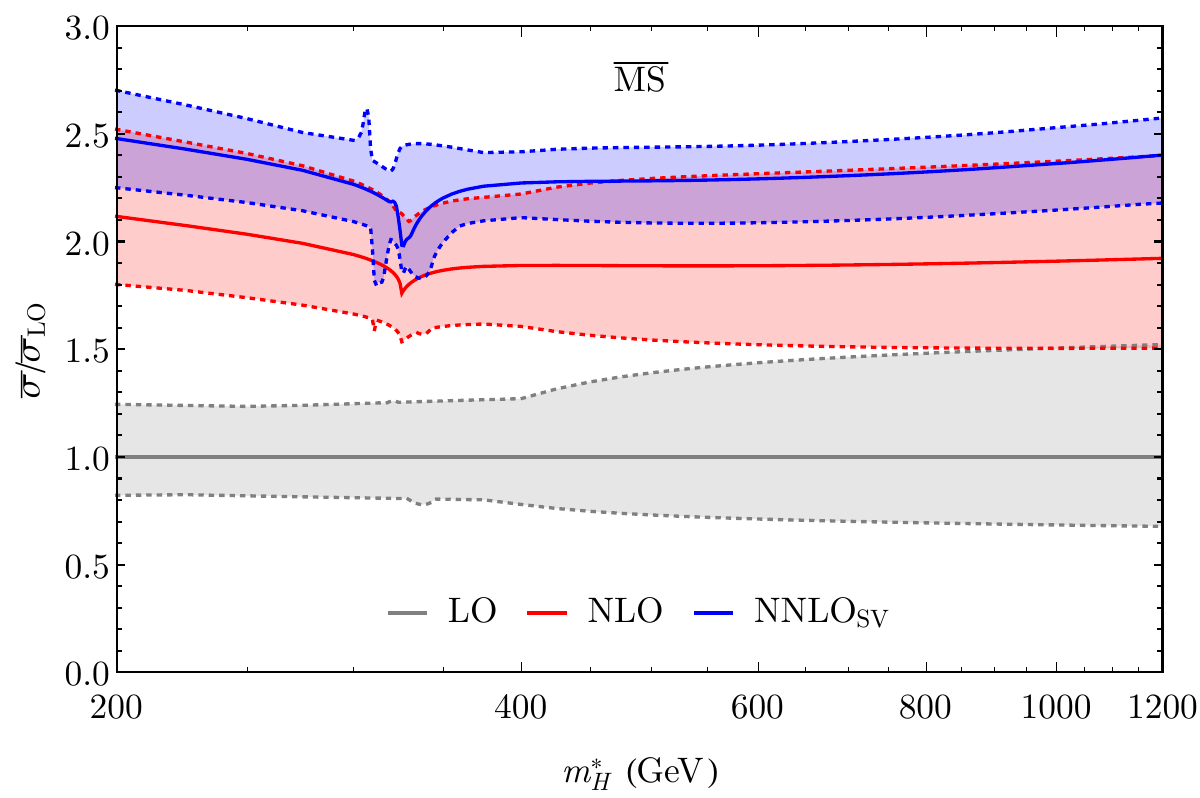}
\includegraphics[width=0.49\textwidth]{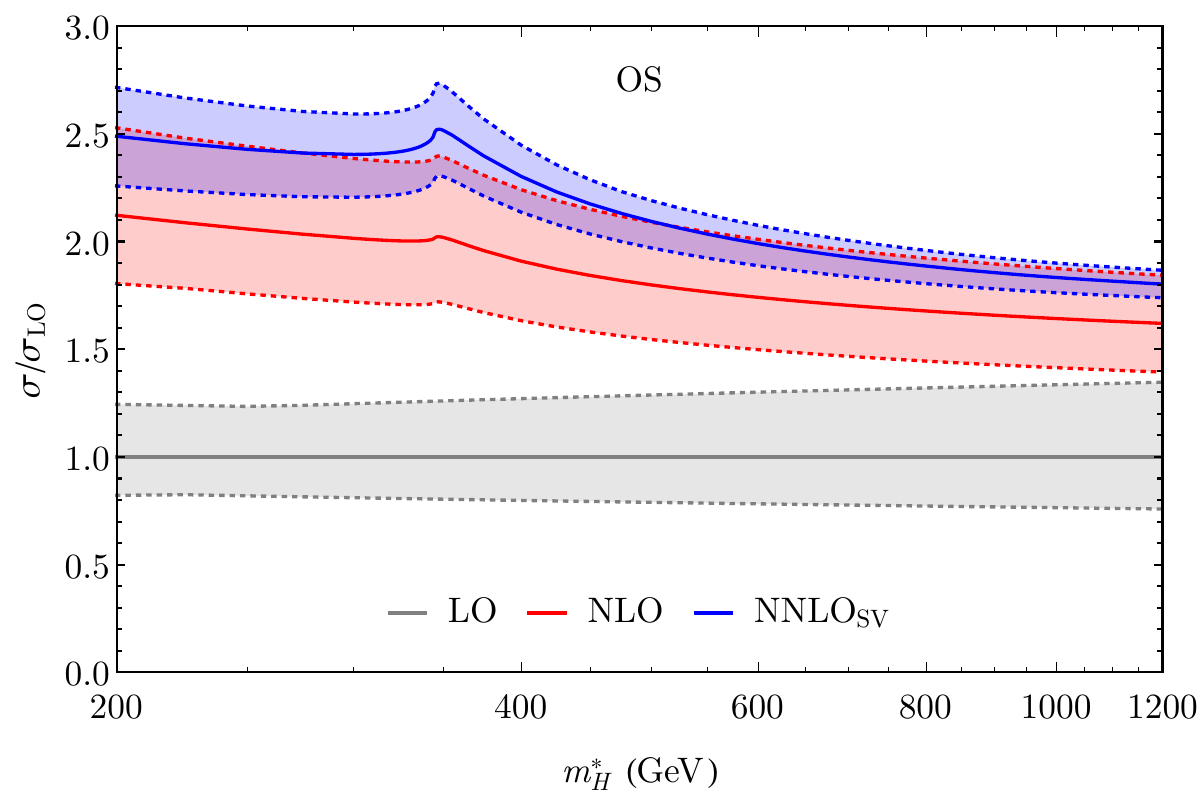}
\end{center}
\vspace{-4ex}
\caption{\label{fig:Kfac}\small
$K$-factors in the \ms (left) and OS (right) schemes, as a function of $m_H^*$. Predictions at LO (gray), NLO (red) and \nnlosv (blue) are shown, together with the corresponding uncertainty bands. The central scale $\mu_0=m_H^*/2$ is always used in the denominator.
}
\end{figure}

We now turn our attention to the large invariant mass region. 
As can be observed from figure~\ref{fig:MS_over_OS_bands}, there is an increasing difference in the behaviour of the \OS and \ms schemes with $m_H^*$, though the uncertainty bands of the two predictions always overlap.
This overlap is only marginal at \nnlosv, partially due to the fact that the \OS prediction presents a very strong reduction of the scale uncertainties at this order.\footnote{We note that this strong reduction at large $m_H^*$, of almost a factor of 4 w.r.t. the NLO uncertainty, might point to an underestimation of the true perturbative uncertainties at \nnlosv for this particular scale choice.}
The difference in the behaviour of the \OS and \ms results is associated to the numerical difference between $M_t$ and $m_t(\mum)$, which also grows as \mum becomes larger.
In order to understand the convergence of the perturbative series in both schemes, it is illustrative to look at the corresponding $K$-factors.
This is presented in figure~\ref{fig:Kfac}.

From the comparison of the two sets of $K$-factors in figure~\ref{fig:Kfac}, we can observe that the \OS scheme presents a better convergence in the $m_H^*>500$~GeV region. The QCD corrections are significantly smaller than their \ms counterparts, for instance increasing the cross section by 62\% at NLO and a further 11\% at \nnlosv for $m_H^*=1200$~GeV, while the corresponding \ms results are 92\% and 25\%, respectively.
Assuming that further QCD corrections follow a similar trend, we expect the \OS prediction at \nnlosv to suffer from small higher-order corrections, while the ones in the \ms scheme are expected to be more sizeable. This fact, combined with the observation that \OS and \ms predictions come closer to each other as we increase the order of the calculation, suggests that the \ms prediction will present larger perturbative corrections to eventually converge to the \OS result, the latter being more stable upon the effect of QCD corrections.

From the behaviour of the perturbative expansion described in the previous paragraph, we conclude that the \OS prediction seems to be the preferred one at large $m_H^*$. This conclusion is based on the comparison of the phenomenological results obtained in the \ms and \OS schemes, and on the assumption that the features observed up to NNLO extend to higher orders. A complementary study on the analytic structure of the cross section in both schemes, and in particular the structure of logarithmically-enhanced terms in this kinematical region at NNLO and beyond (along the lines of refs.~\cite{Liu:2017vkm,Liu:2018czl,Anastasiou:2020vkr}), might provide additional tools to decide the preferred scheme and scale choice at large values of $m_H^*$.

We finally comment on the way the results in the different schemes can be combined in order to provide an estimation of the theoretical uncertainties. The most conservative approach would correspond to taking the envelope of the \OS and \ms predictions, where each of them is computed using 7-point and 15-point variations, respectively (i.e., using independent variations for all the scales involved). In this way, we would be considering at the same time the ``usual" \muR and \muF uncertainties and the ones arising from the scheme choice, and also from the scale choice \mum. An alternative approach, used for instance in ref.~\cite{Baglio:2020wgt} in the context of di-Higgs production, can be defined by linearly adding the uncertainty coming from \mum variation (defined relative to the \OS prediction) at fixed $\muR=\muF=\mu_0$ values to the usual 7-point variation obtained in the \OS scheme. We have checked that both approaches lead to quantitatively similar results.
We point out, however, that either of these approaches can be deemed conservative if there are reasons to prefer one scheme over the other, as suggested by our analysis in the large $m_H^*$ region.

Before presenting our concluding remarks, we can attempt to provide an insight on the more complicated process of Higgs boson pair production. Since the corresponding three-loop corrections with full $M_t$ dependence are currently unknown, we cannot perform the same type of analysis done in the case of off-shell Higgs boson production.
The renormalization-scheme uncertainties have been studied up to NLO in refs.~\cite{Baglio:2018lrj,Baglio:2020ini} and, even though they were found to be larger than the corresponding single-Higgs case, we can observe that they show a similar pattern. In the absence of a better way to estimate these uncertainties, we can use the results presented below as a rough estimate of the expected behaviour at NNLO.

In ref.~\cite{Baglio:2020ini} an uncertainty to the total di-Higgs production cross section was provided by comparing the \OS results (used as a central prediction) to the ones in the \ms scheme. The procedure consisted in independently summing the uncertainties in each $m_{HH}$ bin in order to obtain the total uncertainty. The uncertainty in each bin was estimated varying $\mum$ for fixed values $\muR=\muF=m_{HH}/2$. 

Following a similar approach, we can use the ratio $\bar{\sigma}/\sigma$ shown in figure~\ref{fig:MS_over_OS_bands} as an estimate of the corresponding di-Higgs quantity $(d\bar{\sigma}^{HH}/dm_{HH})/(d\sigma^{HH}/dm_{HH})$, where the invariant mass $m_H^*$ is identified with the invariant mass of the di-Higgs system, $m_{HH}$.
Note that this approach is exact for the di-Higgs contributions arising from a triangle loop, while in practice, however, the box contribution is dominant, especially at large invariant masses.

We therefore proceed as follows: we multiply the di-Higgs $m_{HH}$ distribution computed in the \OS scheme by the LO, NLO and \nnlosv curves defined by the envelopes of the bands present in figure~\ref{fig:MS_over_OS_bands}, that is, the envelope of both \OS and \ms predictions.\footnote{For the LO, NLO and \nnlosv uncertainty estimates, the \OS-scheme di-Higgs $m_{HH}$ distribution is computed at LO, NLO~\cite{Borowka:2016ehy,Baglio:2018lrj} and NNLO$_{\rm FTapprox}$~\cite{Grazzini:2018bsd}, respectively. We note, however, that only the shape of the distribution is relevant for our purposes, and for instance the numbers quoted in eq.~(\ref{eq:HH}) change only at the per-mille level if the LO $m_{HH}$ spectrum is used for the NLO and \nnlosv uncertainty estimates.} We afterwards integrate over $m_{HH}$ to obtain a total cross section, and compare the result to the original prediction in the \OS scheme to estimate the uncertainty. Note that the approach of using the envelope of the invariant mass distribution leads to larger effects than those that would be obtained by computing the total cross section for each scale setting independently, since the scheme uncertainties would be reduced due to a partial cancellation between the effects in the low and high invariant mass regions.
As mentioned before, taking the envelope of the \OS and \ms bands leads to results numerically close to those obtained by linearly adding the \OS scale uncertainties with those coming from exclusive $\mum$ variation. 

Based on the approach described above, we obtain the following results:
\begin{eqnarray}
\label{eq:HH}
\Delta \sigma_{\rm LO}^{HH} &\sim& +33\% -30\% \; (+6.3\% -11\%) \;, \nn\\
\Delta \sigma_{\rm NLO}^{HH} &\sim& +19\% -20\% \; (+3.9\% -6.3\%) \;, \\
\Delta \sigma_{\rm NNLO}^{HH} &\sim& +7.7\% -10\% \; (+2.2\% -3.1\%) \;, \nn
\end{eqnarray}
where the first two numbers of each line correspond to the combined scheme and scale uncertainty estimated as explained in the previous paragraph, while the numbers in parenthesis show the uncertainty coming from only varying $\mum$ at fixed values $\muR=\muF=m_{HH}/2$. 
If we compare our NLO result in eq.~(\ref{eq:HH}) to the NLO uncertainties due to $\mum$ variation reported in ref.~\cite{Baglio:2020ini} ($+4\%-18\%$, see eq.~(4.6) of ref.~\cite{Baglio:2020ini}), we can see that the upper bound is in excellent agreement, while our lower bound underestimates the uncertainties quoted in ref.~\cite{Baglio:2020ini}. This can be understood at LO from the following observations. The box contribution, which is dominant in di-Higgs, has a stronger $M_t$ dependence than the triangle contribution. However, the effect of lowering $M_t$ produces, in both cases, an enhancement in the cross section at low $m_{HH}$, and a suppression in the tail. The crossover between these two regions happens at different values of $m_{HH}$ for the box and the triangle, and in particular it is larger for latter. This fact, therefore, creates a larger positive effect in the triangle that compensates the smaller $M_t$ dependence, while at the same time increases the difference w.r.t. the box in the negative variation. Since our estimates in eq.~(\ref{eq:HH}) are based on approximating the $M_t$ dependence of the full di-Higgs cross section by the one of the triangle contribution, the previous analysis directly translates into our results. 

From the numbers in eq.~(\ref{eq:HH}) we can observe that both the combined uncertainties and the exclusive \mum variation present a significant reduction at \nnlosv, of about a factor of two w.r.t. the NLO predictions. Even if this exercise can only be understood as a rough estimate of the renormalization-scheme ambiguity in di-Higgs production, the reduced uncertainties at \nnlosv observed in this work are an encouraging result. In particular, given the excellent agreement between our NLO estimation for the upper bound in eq.~(\ref{eq:HH}) and the full NLO presented in ref.~\cite{Baglio:2020ini}, the \nnlosv result of $+2.2\%$ in the last line of eq.~(\ref{eq:HH}) can be used as a rough estimate of the upper \mum uncertainty at that perturbative order.

\begin{figure}
\begin{center}
\includegraphics[width=0.49\textwidth]{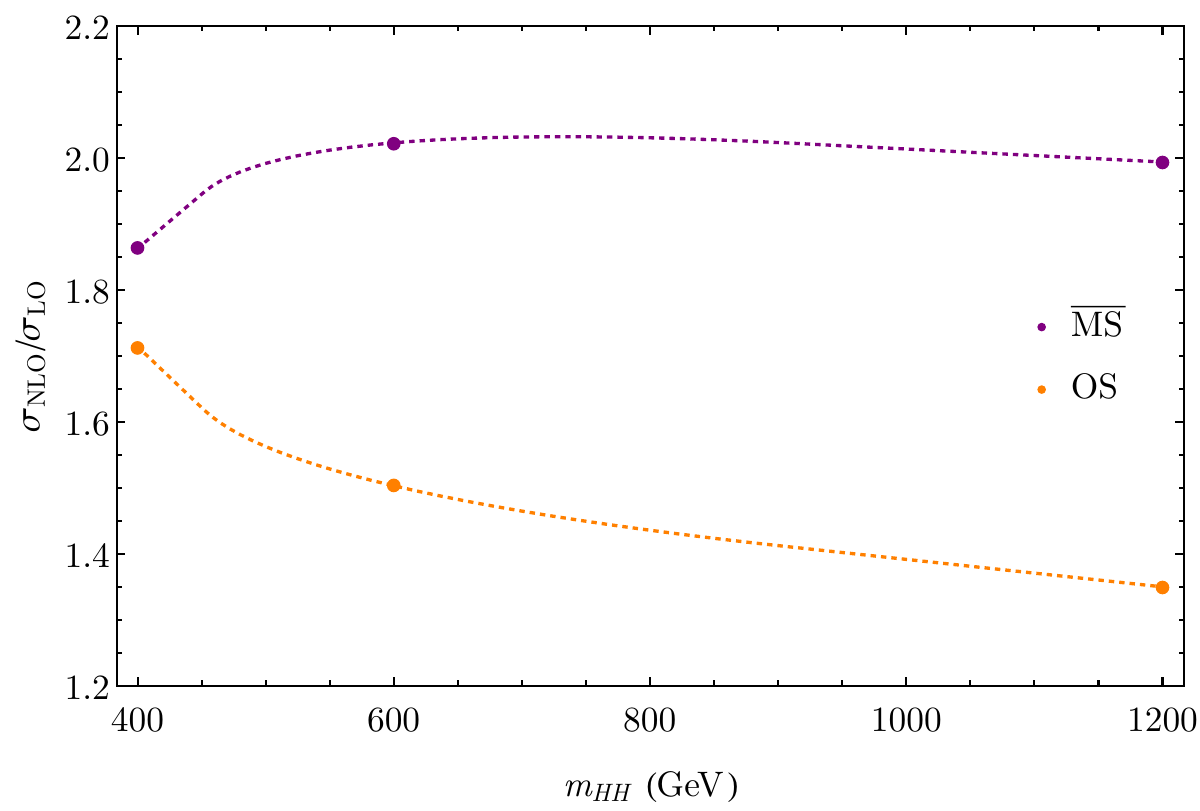}
\end{center}
\vspace{-4ex}
\caption{\label{fig:Kfac_HH}\small
NLO $K$-factors for di-Higgs production as a function of $m_{HH}$ in the \ms (purple) and \OS (orange) schemes. The values, indicated with dots, are taken from ref.~\cite{Baglio:2020ini}, and the interpolating curves are shown only for illustrative purposes.
}
\end{figure}

While the lower bound of the uncertainty also presents a significant reduction in eq.~(\ref{eq:HH}), the large difference w.r.t. the NLO result of ref.~\cite{Baglio:2020ini} makes it harder to justify its use in the context of di-Higgs production.
We can, however, have a look at the corresponding di-Higgs $K$-factors, in a similar way to what we did in figure~\ref{fig:Kfac} for the case of an off-shell Higgs boson.
Since the NLO corrections for $HH$ production with an arbitrary top-quark mass are not publicly available, we rely on the predictions that are present in the literature. The corresponding LO and NLO cross sections for $m_{HH}=400$~GeV, 600~GeV and 1200~GeV are then taken from ref.~\cite{Baglio:2020ini}, and the $K$-factors in both renormalization schemes are presented in figure~\ref{fig:Kfac_HH}.

We can observe that the results shown in figure~\ref{fig:Kfac_HH} present a very similar pattern to those in figure~\ref{fig:Kfac}, i.e. the \OS scheme has a better perturbative behaviour than the \ms scheme in the large invariant mass region. We can therefore expect again that the \OS prediction will be more stable from the perturbative point of view, and that larger higher-order corrections to the \ms prediction will bring it closer to the former.
In light of this, considering the envelope of the \OS and \ms predictions in the tail of the $m_{HH}$ distribution, and the ensuing effect in the total cross section, is a rather conservative approach, and a more aggressive alternative (like considering only the scale uncertainties in the \OS scheme in this invariant mass region) might be considered. 
Since in the large invariant mass region we observe the largest suppression in the \ms predictions compared to the \OS ones, such approach would lead to a reduced estimate of the lower bound of the scheme and scale uncertainties.

It is worth pointing out that in ref.~\cite{Baglio:2020ini}, and in contrast to our conclusions based on the perturbative convergence of the different renormalization schemes, it was shown that the logarithmic structure of the NLO corrections in the high-energy limit~\cite{Davies:2018qvx} suggests that a running top-quark mass is preferred for large values of $m_{HH}$.\footnote{We note that this observation only applies to the case of di-Higgs production, for which the LO squared amplitude does not contain logarithms in the high-energy limit. This is not the case for off-shell single Higgs production, nor for other top-quark-induced processes like $ZH$ production~\cite{Chen:2022rua}.}
In view of this, higher-order studies on the logarithmic structure of the di-Higgs cross section in this kinematical region, as well as results including the top-quark mass dependence in the three-loop virtual corrections (or, eventually, approximate results valid at large $m_{HH}$), are definitely desirable, and might shed light on the preferred scheme and scale choice in the tail of the invariant mass distribution.

Finally, we note that in the limit in which the Higgs self coupling $\lambda$ becomes very large the triangle contributions dominate the di-Higgs cross section, and therefore we can apply our results for off-shell Higgs boson production without performing any approximation. In that case, we find the following uncertainties:
\begin{eqnarray}
\label{eq:HHlargeLambda}
\Delta \sigma_{\rm LO}^{HH}(\lambda\to\infty) &=& +36\% -22\% \; (+17\% -3.6\%) \;, \nn\\
\Delta \sigma_{\rm NLO}^{HH}(\lambda\to\infty) &=& +23\% -16\% \; (+8.3\% -1.8\%) \;, \\
\Delta \sigma_{\nnlosv}^{HH}(\lambda\to\infty) &=& +12\% -9.0\% \; (+4.9\% -1.3\%) \;, \nn
\end{eqnarray}
where the notation is the same as in eq.~(\ref{eq:HH}). We can observe that the $\mum$ variation has a similar overall magnitude as for $\lambda=1$, though in this case the upper uncertainty is significantly larger than the lower one. This is expected, since the triangle contribution is concentrated at lower values of $m_{HH}$ (the peak, for instance, is located around $m_{HH}\sim 270$~GeV, to be compared to  $m_{HH}\sim 400$~GeV in the $\lambda=1$ case), and this in turn is the region where the \ms prediction presents larger positive deviations w.r.t. the \OS result.

\section{Conclusions}
\label{sec:conclusions}

In this work we have performed a NNLO study of the uncertainties related to the renormalization scheme of the top-quark mass in Higgs boson production via gluon fusion.
We have focused on the case of off-shell production, where the large value of $m_H^*$ leads to a stronger parametric dependence on the top-quark mass, which in turn generates much larger renormalization scheme uncertainties compared to the on-shell case.

Based on the three-loop results with full top-quark mass dependence presented in ref.~\cite{Czakon:2020vql}, and profiting from the known structure of QCD cross sections in the soft limit~\cite{deFlorian:2012za}, we have constructed the soft-virtual approximation for Higgs boson production at NNLO. Our \nnlosv predictions therefore contain the full top-quark mass dependence, with the only caveat that NNLO real corrections are treated in the soft limit. We have also evaluated the \nlosv cross section, and found excellent agreement with the full NLO result, therefore validating our approach at the next perturbative order.

We have obtained \nnlosv predictions using two different schemes for the renormalization of the top-quark mass: the on-shell scheme and the \ms scheme. In the latter, an additional renormalization scale \mum is introduced. We have evaluated the scale uncertainties by allowing independent variations of all the relevant scales, leading to the usual 7-point variation in the \OS scheme, and a 15-point variation in the \ms scheme.

We have found sizeable differences between the predictions obtained in the two schemes, both close to the $t\bar{t}$ threshold and in the large $m_H^*$ region. While the differences are large, the results are always compatible within scale uncertainties. We have also observed a significant reduction in the difference between the two schemes as higher orders in the QCD perturbative expansion are included.
We have studied the quality of the perturbative convergence in both schemes by computing the $K$-factors, and found that for large values of $m_H^*$, that is $m_H^*>500$~GeV, the \ms results present much larger corrections than their \OS counterparts.
We concluded that the \OS scheme is the preferred choice in this invariant mass region, though further studies addressing the logarithmic structure of the cross section in this limit would be desirable.

Using the results obtained for off-shell Higgs boson production, we have derived a rough estimate of the renormalization scheme and scale uncertainties for di-Higgs production. Our approximation of the upper uncertainty band is in line with the actual uncertainties at NLO, and our results suggest that it would be substantially reduced at NNLO.
The lower band is driven by the large $m_{HH}$ region, where the analysis of the $K$-factors in off-shell Higgs production at \nnlosv, as well as in di-Higgs production up to NLO, shows that the \OS predictions have a much better perturbative convergence. 
A more detailed study for di-Higgs production, which is beyond the scope of this paper, should consistently account for the box-diagram contributions.

\section*{Acknowledgements}

The author would like to thank Massimiliano Grazzini and Michael Spira for useful discussions and comments on the manuscript.

\bibliography{biblio}

\end{document}